\documentclass[twocolumn,10pt]{article}
\usepackage{graphics}

\begin{document}
\title{Dynamics and thermodynamics in spinor quantum gases}
\author{\small H. Schmaljohann, M. Erhard, J. Kronj{\"a}ger, K. Sengstock and K. Bongs\\
\small Institut f\"ur Laser-Physik, Universit\"at Hamburg, Luruper
Chaussee 149, 22761 Hamburg, Germany}

\maketitle
\begin{abstract}
We discuss magnetism in spinor quantum gases theoretically and
experimentally with emphasis on temporal dynamics of the spinor
order parameter in the presence of an external magnetic field. In
a simple coupled Gross-Pitaevskii picture we observe a dramatic
suppression of spin dynamics due to quadratic Zeeman
''dephasing''. In view of an inhomogeneous density profile of the
trapped condensate we present evidence of spatial variations of
spin dynamics. In addition we study spinor quantum gases as a
model system for thermodynamics of Bose-Einstein condensation. As
a particular example we present measurements on condensate
magnetisation due to the interaction with a thermal bath.
\end{abstract}
\section{Introduction}\label{intro}
The field of cold quantum gases has seen a rapid growth since the
first realisation of Bose-Einstein condensation in dilute atomic
gases in 1995~\cite{Anderson1995a,Davis1995b,Bradley1995a}
accompanied with the development of a broad range of tools for the
detailed control of these systems. Single component Bose-Einstein
condensates have evolved into a fundamental model system showing
many intriguing phenomena. For an overview see
e.g.~\cite{Cornell1999a,Ketterle1999a,Dalfovo1999a,Courteille2001a,Pethick2002a,Pitaevskii2003a,Andersen2003a,Bongs2004a}.
In contrast to these systems with a scalar order parameter spinor
Bose-Einstein condensates offer spin as a new degree of freedom
and are consequently represented by a vector order parameter. In
addition to being mixtures of different bosonic species the
different components in these multicomponent quantum systems are
coupled and can exchange particles. This makes spinor
Bose-Einstein condensates unique systems, which on the one hand
possess intrinsic magnetic properties and on the other hand give
access to well controlled Bose-Einstein thermodynamics with
adjustable heat and particle bath.

Magnetism in degenerate quantum gases offers new regimes for
studies of collective spin
phenomena~\cite{Ho1998a,Ohmi1998a,Isoshima1999c,Koashi2000a,Ho2000a,Ho2000b,Ciobanu2000a,Myatt1997a,Hall1998a,Matthews1998a,Stenger1999a,Miesner1999a,Stamper-Kurn1999b,McGuirk2002a,Leanhardt2003b,Schmaljohann2004a,Chang2004a}
and opens new perspectives in view of the closely related
entangled spin systems in atomic quantum gases, which show
intriguing prospects for quantum optics and quantum
computation~\cite{Pu2000a,You2000a,Soerensen2001a,Julsgaard2001a,Mandel2003a}.

So far studies concentrated on the magnetic properties of spin 1
ultracold quantum gases in optically trapped
$^{23}$Na~\cite{Stenger1999a,Miesner1999a,Stamper-Kurn1999b,Leanhardt2003b}
and recently in
$^{87}$Rb~\cite{Barrett2001a,Schmaljohann2004a,Chang2004a,Kuwamoto2004a},
where also the intrinsically more complex F=2 spin state became
accessible~\cite{Schmaljohann2004a,Kuwamoto2004a}.

Systems closely related to these spinor condensates are effective
spin-1/2 systems realized by radiofrequency coupling of two
hyperfine states in
$^{87}$Rb~\cite{Myatt1997a,Hall1998a,Matthews1998a}, in which
spin-waves~\cite{McGuirk2002a} and decoherence effects were
observed~\cite{Lewandowski2003a}.

In this paper we will concentrate on F=1 and F=2 spinor
condensates of $^{87}$Rb in two limits. First we investigate the
coherent spinor evolution of a trapped ensemble at zero
temperature in the presence of a homogeneous magnetic field, where
we find suppression of spin dynamics due to the quadratic Zeeman
effect as well as a spatial depedence of the dynamics. The other
limit is thermally dominated spin dynamics at temperatures close
to $T_c$, where a significant fraction of the atoms occupies the
normal component. Spinor gases in this regime can act as a
versatile model system for thermodynamics with tunable heat and
particle bath as recently demonstrated with a constant temperature
Bose-Einstein phase transition~\cite{Erhard2004a}. In this paper
we will present new data on thermally induced condensate
magnetisation as another intriguing example of spinor
thermodynamics.

\section{Spinor condensates at T=0}\label{sec:1}
The theory presented in this section is based on a mean field
approach, in extension of the very successful treatment of single
component Bose-Einstein condensates. The basic two-particle
interactions are represented by a density and spin-composition
dependent average energy shift. This approach has first been
developed for F=1 systems~\cite{Ho1998a,Ohmi1998a} and was later
extended to F=2 systems~\cite{Koashi2000a,Ciobanu2000a}.

For typical experimental parameters the mean field shifts
connected to collisions in different spin channels dominate
magnetic dipole dipole interactions by at least one order of
magnitude. In the following analysis magnetic dipole dipole
interactions will thus be neglected. The intrinsic dynamics of a
spinor condensate is determined by a pairwise interaction
potential~\cite{Ho1998a,Koashi2000a}:
 \begin{equation}
    \hat V(\mathbf{r_1-r_2})=\delta(\mathbf{ r_1-r_2})\sum_{f=0}^{2F}\frac{4\pi
    \hbar^2a_f}{m}\hat P_f.
 \end{equation}
Here $a_f$ denotes the s-wave scattering length for a collision
channel of two particles whose single spins $F$ are combined to
give the total spin $f$, $\hat P_f$ is the corresponding
projection operator onto total spin $f$ and $m$ is the mass of a
single atom. Due to Bose symmetry only even total spin channels
(e.g. $a_0,a_2,a_4$ for $F=2$) are involved with the maximum total
spin given by $f=2F$. Making use of the relation $\left( \vec F_1
\cdot \vec F_2 \right)^n = \sum_{f=0}^{2F} \lambda_f^n P_f$ with
$\lambda_f = \frac{1}{2} \left[ f(f+1)-2F(F+1)\right] $ the
projection operators can be replaced by spin expectation
values~\cite{Ho1998a}.

In the following we will concentrate on the case $F=1$ for the
theoretical considerations in order to point out some important
aspects of spin dynamics, which are straightforward to extend to
the $F=2$ case.

In second quantised form the Hamiltonian for a $F=1$ system at
zero magnetic field is given by~\cite{Ho1998a}:
\begin{eqnarray}
  H=\int d^3 r \left( \frac{\hbar^2}{2m} \nabla \psi_a^{\dagger
  }\cdot \nabla \psi_a +V_{ext} \psi_a^{\dagger}\psi_a \right. \nonumber \\
  +
  \frac{g_0}{2}\psi_a^{\dagger}\psi_{a'}^{\dagger}\psi_{a'}\psi_a  \nonumber \\
  \left. + \frac{g_2}{2}\psi_a^{\dagger}\psi_{a'}^{\dagger}\mathbf{\vec F_{ab}\cdot
  \vec F_{a'b'}}\psi_{b'}\psi_b\right) .
\end{eqnarray}
In this expression $\psi_a(\vec r)$ is the field annihiliation
operator for an atom in state $m_F=a$ at point $\vec r$ and
$V_{ext}$ is the trapping potential. The spin-independent
mean-field interaction is parameterised by $g_0=\frac{2\pi
\hbar^2}{m}\times \frac{2a_2+a_0}{3}$. The spin-dependent
mean-field responsible for the systems magnetic properties is
characterised by the parameter
$g_2=\frac{4\pi\hbar^2}{m}\times\frac{a_2-a_0}{3}$ and the
coupling between different states is determined by the
spin-matrices:
\begin{eqnarray}
   F_x=\frac{1}{\sqrt{2}}\left(%
\begin{array}{ccc}
  0 & 1 & 0 \\
  1 & 0 & 1 \\
  0 & 1 & 0 \\
\end{array}%
\right) ,\quad
   F_y=\frac{i}{\sqrt{2}}\left(%
\begin{array}{ccc}
  0 & -1 & 0 \\
  1 & 0 & -1 \\
  0 & 1 & 0 \\
\end{array}%
\right) ,\\
F_z=\left(%
\begin{array}{ccc}
  1 & 0 & 0 \\
  0 & 0 & 0 \\
  0 & 0 & -1 \\
\end{array}%
\right)
\end{eqnarray}
From the spin matrices it follows, that for F=1 spinor condensates
the only coupling process is between states with $m_F=0$ and
$m_F=\pm 1$ and that the total spin projection is preserved by the
interaction. It is important to emphasise this again to
demonstrate that the total spin in a finite atomic quantum gas
system is conserved, in contrast to a homogeneous infinite system
and in contrast to many condensed matter systems. Therefore the
actual ground state of the system depends on the initial
magnetisation.

For macroscopically occupied Bose-systems at $T=0$ it is common to
replace the field annihilation operators by their expectation
value, i.e. $\varphi_a(\vec r,t) \equiv \langle \psi_a (\vec
r,t)\rangle $, which for spinor condensates is conveniently
expressed as~\cite{Ho1998a}
\begin{equation}\label{eq:1}
  \varphi_a(\vec r,t) = \sqrt{n(\vec r,t)}e^{i\phi (\vec r,t)}\zeta_a (\vec r,t) .
\end{equation}
Here $n(\vec r,t)$ is the condensate density, $\phi (\vec r,t) $ a
phase and $\vec \zeta (\vec r,t) = (\zeta_{+1}, \zeta_0,
\zeta_{-1})^T$ is a normalized spinor with $\vec \zeta^{\dag }
\cdot \vec \zeta =1$.

Using (\ref{eq:1}), neglecting the density dependence on the spin
state and adding the effect of a weak magnetic field one gets the
following system of differential equations for the evolution of an
F=1 spinor condensate:
\begin{eqnarray}
i\hbar\frac{\partial}{\partial t}\sqrt{n(\vec{r},t)}e^{i\phi(\vec r, t)}&=&\left(-\frac{\hbar^2\nabla^2}{2m}+V_{ext.}(\vec{r})+g_0n(\vec{r})\right)\nonumber \\
 & & \cdot \sqrt{n(\vec{r},t)}e^{i\phi(\vec r,t)}\,, \label{gleichungdichteentwicklung}\\
i\hbar\frac{\partial}{\partial t}\vec{\zeta}(\vec{r},t)&=&-\frac{\hbar^2n(\vec{r},t)\nabla^2}{2m}\,\vec{\zeta}(\vec{r},t) \nonumber \\
 & & +g_2n(\vec{r},t)\mathcal{\vec{F}}\vec{\zeta}(\vec{r},t)\vec{\zeta}^{\dag }(\vec{r},t)\mathcal{\vec{F}}\vec{\zeta}(\vec{r},t)\nonumber\\
&&-p\mathcal{F}_z\vec{\zeta}(\vec{r},t)+q(\mathcal{F}_z^2-4)\vec{\zeta}(\vec{r},t)\,.
\label{eq:spindynamik}
\end{eqnarray}
From these equations it follows, that spin dynamics (represented
by the terms containing $\mathcal{\vec{F}}$) is proportional to
$g_2n(\vec r,t)$, i.e. the local density. This has important
consequences for the usual experimental situation of trapped
samples, having an inhomogeneous density distribution. Two regimes
can be identified: 1. the spin dynamics rates are higher than one
or more trapping frequencies and 2. spin dynamics is slow compared
to trap dynamics. In the first case there will be a significant
coupling between spin dynamics and motional dynamics, while the
second case is comparable to the homogeneous density case. As was
shown for the first time in~\cite{Schmaljohann2004a} $^{87}$Rb has
the fascinating properties that it offers fast spin dynamics in
the F=2 state (on the order of a few ms) and slow spin dynamics in
the F=1 state (on the order of a few s). Therefore also different
spatial regimes can be thought of with $^{87}$Rb spinor
condensates.

\begin{figure}
\resizebox{0.45\textwidth}{!}{
  \includegraphics{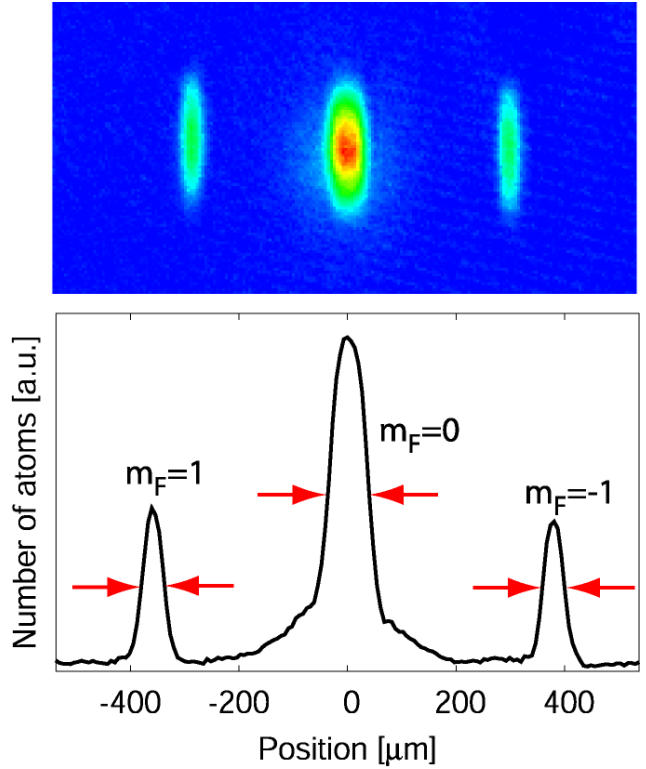}
}
 \caption{Absorption images and vertically summed cross sections
           of a spinor condensate after 10\,ms evolution starting
           with the initially prepared state $F=2,\, m_F=0$. The images
           were taken after a combined time of flight
           and Stern Gerlach separation, leading to a rapid radial expansion and
           a separation of the different spinor components. The graphs clearly show
           a reduced width in the axial distribution of the $m_F=\pm 1$ components
           created by spin dynamics in the high density center of the $m_F=0$ component.}\label{f:spatialdynamics}
\end{figure}

Fig.~\ref{f:spatialdynamics} shows the initial evolution of an F=2
$^{87}$Rb spinor condensate prepared in the $m_F=0$ spin state
with a central density $n\approx 4\cdot 10^{14}\, $cm$^{-3}$. In
this case spin dynamics takes place with timescales on the order
of a few ms~\cite{Schmaljohann2004a}, faster than the axial trap
dynamics characterised by a frequency of $\approx 21\, $Hz. The
other trapping frequencies of $\approx 155\,$Hz and $\approx 890\,
$Hz imply motional dynamics faster than the spin evolution and are
thus neglected in the following. The image in
Fig.~\ref{f:spatialdynamics} was taken after switching off the
trapping potential and 31\,ms time of flight during which a
magnetic field gradient was applied for 5\,ms to achieve a
Stern-Gerlach separation of the spinor components. The axial trap
direction was horizontal, along which only weak mean field induced
expansion takes place during time of flight. The vertically summed
horizontal cross sections shown in Fig.~\ref{f:spatialdynamics}
thus approximately represent the axial distribution of the trapped
sample. The images and graphs clearly demonstrate the density
dependence of spin dynamics, as the $m_F=\pm 1$ spin states show a
smaller width than the ''mother'' $m_F=0$ component. This is due
to the parabolic density profile of the ''mother'' component,
which implies a fast spin state conversion in the center of the
trap. Very recently a spatial dependence in $^{87}$Rb (F=2)
condensate spin dynamics was also observed
in~\cite{Kuwamoto2004a}. Further investigations of the coupled
spin and spatial dynamics were limited in our case by the finite
optical resolution and will be subject to future investigations.

Interestingly external magnetic fields are a further additional
parameter to control spin dynamics. Whereas most theoretical work
so far concentrated on the physics at B=0, experiments
~\cite{Stamper-Kurn1999b,Schmaljohann2004a,Chang2004a,Kuwamoto2004a}
clearly demonstrated the importance of extermal field influences.
Magnetic fields can completely hinder spin dynamics or on the
other hand strongly stimulate dynamics. In the following we will
mainly concentrate on a discussion of spin dynamics suppression
due to the quadratic Zeeman effect. Very recently the suppression
of spin dynamics in F=2 spinor condensates was experimentally
observed in~\cite{Kuwamoto2004a}. The following discussion can
explain these observations in their main part.

The magnetic field enters equations (\ref{eq:spindynamik}) via the
linear Zeeman shift as well as the quadratic Zeeman shift with the
coupling constants $p= g_F\mu_B B$ and
$q=-\frac{\mu_B^2B^2}{4\hbar \omega_{12}}$ with $\omega_{12}$
representing the hyperfine splitting. As we will see later the
relatively large linear Zeeman energy ($q/B \approx k_B\cdot 34\mu
$K/G) does not influence the spin dynamics, which in fact is due
to total spin conservation. This is not true for the quadratic
Zeeman effect! For typical experimental conditions with offset
fields of several 100$\, $mG the quadratic Zeeman energy ($|q|/B^2
\approx k_B \cdot 3.5\, $nK/G$^2$) can reach values comparable to
the intrinsic spin coupling (on the order of $k_B$ times one nK
for typical condensate densities of a few $10^{14}\, $cm$^{-3}$).

In order to extract the basic influence of the quadratic Zeeman
effect on spin dynamics, we assume a homogeneous case with
constant $n(\vec r,t)=n$ and no spatial spin variation $\vec \zeta
(\vec r,t) = \vec \zeta(t)$. With these assumptions equation
(\ref{eq:spindynamik}) reads
\begin{eqnarray}\label{eq:spinonedynamic}
i\hbar\frac{\partial}{\partial t}\zeta_{+1}&=&g_2n\,(\zeta^*_{+1}\zeta_{+1}\zeta_{+1}+\zeta^*_0\zeta_{+1}\zeta_0-\zeta^*_{-1}\zeta_{+1}\zeta_{-1}\nonumber \\
     && +\zeta_{-1}^*\zeta^2_0)-p\zeta_{+1}-3q\zeta_{+1}\,, \nonumber \\
i\hbar\frac{\partial}{\partial t}\zeta_0&=&g_2n\,(\zeta^*_{+1}\zeta_{+1}\zeta_0+\zeta^*_{-1}\zeta_0\zeta_{-1}+2\zeta_0^*\zeta_{+1}\zeta_{-1})\nonumber \\
 &&-4q\zeta_0\,,\\
i\hbar\frac{\partial}{\partial
t}\zeta_{-1}&=&g_2n\,(\zeta^*_{-1}\zeta_{-1}\zeta_{-1}-\zeta^*_{+1}\zeta_{+1}\zeta_{-1}+\zeta^*_0\zeta_0\zeta_{-1}\nonumber
\\ && +\zeta_{+1}^*\zeta^2_0)+p\zeta_{-1}-3q\zeta_{-1}\,.\nonumber
\end{eqnarray}

Using a simple change of variables
\begin{eqnarray}
\lambda_{\pm 1}&=&\zeta_{\pm 1}\,\exp(\mp i(p\pm
3q)t/\hbar)\nonumber
\\ &\mbox{and}& \\
\lambda_0&=&\zeta_0\,\exp(-i4qt/\hbar)\nonumber \,,
\end{eqnarray}
the linear Zeeman dependence is removed from the equations and the
quadratic Zeeman effect enters in a more symmetric way:
\begin{eqnarray}\label{eq:nolinearZeeman}
i\hbar\frac{\partial}{\partial t}\lambda_{+1}&=&g_2n\,(\lambda^*_{+1}\lambda_{+1}\lambda_{+1}+\lambda^*_0\lambda_{+1}\lambda_0-\lambda^*_{-1}\lambda_{+1}\lambda_{-1}\nonumber \\
&&+\lambda_{-1}^*\lambda^2_0e^{i2qt/\hbar })\,, \nonumber \\
i\hbar\frac{\partial}{\partial t}\lambda_0&=&g_2n\,(\lambda^*_{+1}\lambda_{+1}\lambda_0+\lambda^*_{-1}\lambda_0\lambda_{-1}\nonumber \\
&&+2\lambda_0^*\lambda_{+1}\lambda_{-1}e^{-i2qt/\hbar })\,,\\
i\hbar\frac{\partial}{\partial
t}\lambda_{-1}&=&g_2n\,(\lambda^*_{-1}\lambda_{-1}\lambda_{-1}-\lambda^*_{+1}\lambda_{+1}\lambda_{-1}+\lambda^*_0\lambda_0\lambda_{-1}\nonumber
\\ && +\lambda_{+1}^*\lambda^2_0e^{i2qt/\hbar })\,.\nonumber
\end{eqnarray}
In these equations derived here spin exchange is described by the
terms with exponentials, while the other terms represent the mean
field phase evolution.
\begin{figure}
\resizebox{0.45\textwidth}{!}{
  \includegraphics{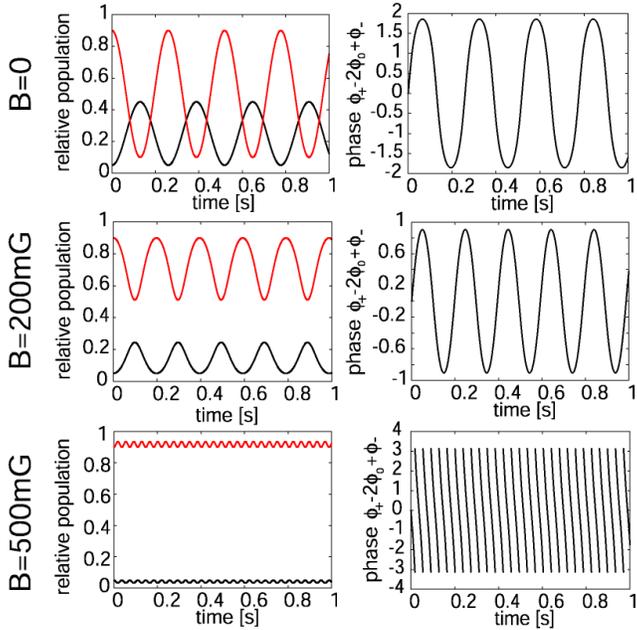}
}
 \caption{Numerical simulation of equations (\ref{eq:nolinearZeeman})
          for different magnetic field strengths. The graphs on the left show
          the evolution of the population of spin state $m_F=0$ (red) and of
          spin states $m_F=\pm 1$ (black). The graphs on the right show the
          evolution of the relative phase corresponding to the
          interaction $2|0\rangle \leftrightarrow |1\rangle +
          |-1\rangle $. A strong reduction of the amplitude of
          the oscillation in the spin populations for higher
          magnetic fields is clearly visible.
           }\label{f:spinsimulation}
\end{figure}
Fig.~\ref{f:spinsimulation} shows the relative spinor occupations
and the relative phase $\phi_{+1} -2 \phi_0 + \phi_{-1}$ as a
result of a numerical simulation of equations
(\ref{eq:nolinearZeeman}) for different magnetic offset fields.
The spin component phases are given by $\phi_i = \arg (\lambda_i)$
with the phase $\phi_{+1} -2 \phi_0 + \phi_{-1}$ is the relevant
phase for the evolution of the $m_F=0$ component. The initial
conditions were chosen symmetric with density $n=4\cdot
10^{14}\,$cm$^{-3}$, $|\lambda_0|=0.9$, $|\lambda_{\pm 1}|=0.05$
and $\phi_i(t=0)=0$.

The numerical simulation clearly shows that the quadratic Zeeman
effect strongly suppresses spin dynamics if it is larger than the
spin dependent mean field shifts. For $^{87}$Rb in F=1 and at
densities of a few times $10^{14}\, $cm$^{-3}$ this suppression
becomes relevant at magnetic fields of a few 100\,mG.

The suppression of spin dynamics at high magnetic fields can also
be directly deduced analysing the expression for $\lambda_0 $,
which according to equation (\ref{eq:nolinearZeeman}) at high
magnetic fields will approximately evolve due to the rapidly
changing exponential giving:
\begin{eqnarray}
\lambda_0(t_{final})& \approx & \lambda_0(t_0)\nonumber
\\ && -\left[\frac{g_2n}{2q}\,\lambda_0^*(t_0)\,\lambda_{-1}(t_0)\,\lambda_{+1}(t_0)\,e^{-i2qt/\hbar}\right]_{t_0}^{t_{final}}\,.
\end{eqnarray}
If $q\gg g_2n$ there will be nearly no change in the occupation of
spin states as predicted by the numerical simulation.

\begin{figure}
\resizebox{0.45\textwidth}{!}{
  \includegraphics{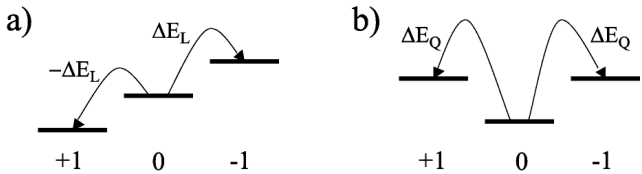}
}
 \caption{Schematic view of F=1 ($|1\rangle +
|-1\rangle \leftrightarrow 2|0\rangle$) spin dynamics with
additional magnetic field. The different $m_F$ states are labeled
by $0$ and $\pm 1$. (a) The total linear Zeeman energy is
conserved in spin dynamics as the energy gain in one component is
lost in the other. (b) The quadratic Zeeman shifts (shown with the
linear Zeeman contribution subtracted and rescaled as typically
$\Delta E_Q \ll \Delta E_L $) however lead to an energy imbalance
in spin dynamics.
           }\label{f:Zeeman}
\end{figure}

In summary (in the absence of a field gradient) the linear Zeeman
effect can be neglected for investigations on spin dynamics in
spinor Bose condensates due to the always symmetric exchange of
Zeeman energy (Fig.~\ref{f:Zeeman}a), which is fundamentally
caused by spin conservation. In contrast we found that spin
dynamics can be significantly altered by the quadratic Zeeman
effect, where an additive energy exchange (Fig.~\ref{f:Zeeman}b)
leads to a ''dephasing'' of the spin components ultimately
stopping spin dynamics at high magnetic fields. One could say that
high external magnetic fields ''pin'' the spin to its value.
Intrinsic spin dynamics can be observed up to magnetic fields for
which $q(B) \simeq g_2n$ typically corresponds to fields of a few
100\,mG. We want to emphasize that the blocking of spin dynamics
is solely due to the quadratic Zeeman effect and does not follow
energetical considerations. Indeed with this blocking effect we
can explain the experimentally observed high magnetic field
suppression of spin dynamics even when it is leading to an
energetically lower state~\cite{Kuwamoto2004a}.

Furthermore we found evidence of spatially varying spin dynamics
in trapped spinor condensates with inhomogeneous density. This
effect will lead to complex coupled dynamics of spatial and spin
degrees of freedom to be investigated in future experiments.

\section{Thermodynamics with spinor condensates} \label{sec:2}
Finite temperature effects in Bose-Einstein condensates represent
an active area of research, which is still relatively unexplored
due to its theoretical complexity and experimental challenges. In
theory sophisticated methods have been developed to reduce the
complexity of simulations such that modelling complex phenomena
seems
feasible~\cite{Sinatra2001a,Jackson2002a,Goral2002a,Morgan2003a,Nikuni2004a}.
Major tests for these models consisted in the interpretation of
early experiments on damping of single component condensate
excitations in the presence of a normal component and on
condensate formation.

\begin{figure}
\resizebox{0.45\textwidth}{!}{
  \includegraphics{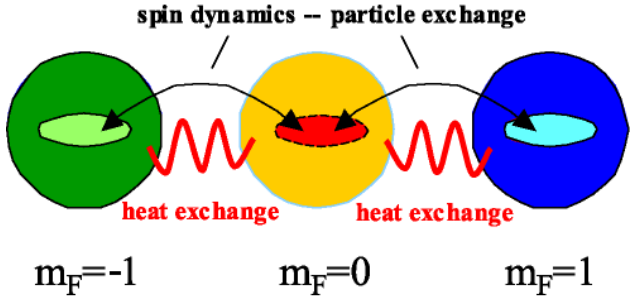}
} \caption{Schematic view of thermodynamics in spinor quantum
gases for the example of a spin F=1 system. In this example the
states with $m_F=1=|1\rangle $ and $m_F=-1=|-1\rangle $ act as a
heat bath and particle reservoir for the state $m_F=0=|0\rangle $.
Heat (energy) is exchanged by elastic collision processes
involving atoms in the normal component. Particles are exchanged
in spin-changing collisions via the interaction $|1\rangle +
|-1\rangle \leftrightarrow 2|0\rangle$. An important aspect of
particle exchange lies in the density dependence of spin dynamics,
which nearly exclusisively takes place in the dense condensate
fraction. Particle exchange thus involves only small energy
transfer and does practically not contribute to thermalization
processes. Spinor condensates thus allow to create systems with
independently tunable particle and heat exchange.}
\label{f:spinorthermodynamics}
\end{figure}

Spinor condensates offer a novel approach to well controlled
Bose-Einstein thermodynamics. As a first aspect they are
multicomponent systems such that a thermal bath for one component
can easily be created by tailoring the other component(s). This
aspect is widely used in sympathetic cooling experiments with
multi-species mixtures. Spinor multicomponent systems add the
essential aspect of particle exchange between the components,
which is required to complete thermodynamics. The particle
exchange takes place due to intrinsic interparticle interactions
but can also be experimentally controlled via additional external
electromagnetic fields. These aspects and the relevant
interactions with respect to thermodynamics in spinor quantum
gases are shown schematically in
Fig.~\ref{f:spinorthermodynamics}.

An important point in the study of spinor systems is connected to
the coherence between different spin components. Coherent spin
mixtures, i.e. mixtures in which each single atom (in the normal
component as well as in the condensate fraction) is in the same
quantum superposition of spin states, are effectively single
component quantum gases, which have to be contrasted to incoherent
spin mixtures, where different spin states represent different
species gases. For example if a F=1 spinor gas in the first case
would be described by $N$ particles in a spin superposition state
$\phi = \alpha |1\rangle + \beta |0\rangle + \gamma |-1\rangle $
then the incoherent state would be given by a mixture of three
gases, one with $N_{1}=|\alpha |^2$ particles in the $|1\rangle $
state, one with $N_{0}=|\beta |^2$ particles in the $|0\rangle $
state and one with $N_{-1}=|\gamma |^2$ particles in the
$|-1\rangle $ state.

Indeed incoherent spin mixtures are in some cases of high
experimental importance, e.g. for the conversion of two spin state
fermion mixtures to a molecular Bose gas with a Feshbach
resonance. This was nicely demonstrated and explained
in~\cite{Strecker2003a} for the preparation of a spin state
mixture in $^6$Li. As an another example the distinction between
coherent and incoherent spin superpositions is crucial to the
understanding of the recently demonstrated decoherence driven
cooling~\cite{Lewandowski2003a} in a quasi spin 1/2 system.

The evolution of F=1 and F=2 spinor condensates discussed in this
paper can be tuned in between the regimes of coherent and
incoherent evolution by adapting the parameters temperature,
density and possibly external radiofrequency coupling. An
intriguing example for mostly coherent evolution is the
observation of spinor
oscillations~\cite{Schmaljohann2004a,Chang2004a,Kuwamoto2004a}.
The incoherent limit was recently reached in a thermalization
dominated regime with the demonstration of constant temperature
Bose-Einstein condensation~\cite{Erhard2004a} in F=1 spinor
condensates with significant occupation in the normal component.

\begin{figure}
\resizebox{0.45\textwidth}{!}{
  \includegraphics{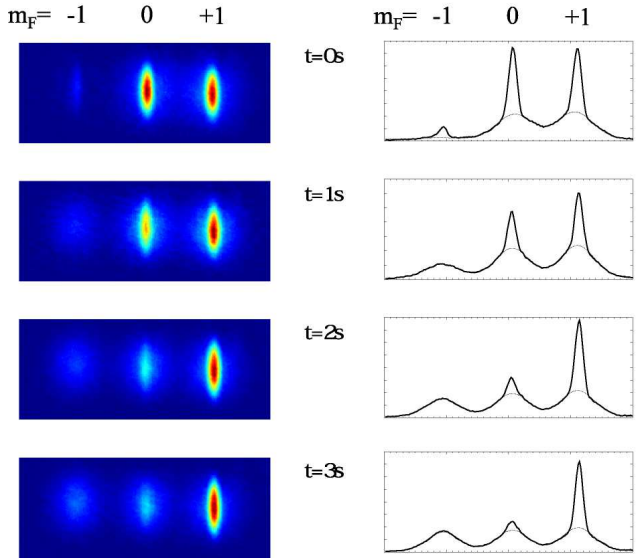}
  }
  \caption{Absorption images and cross sections showing the
  temporal evolution of a spinor condensate prepared in the states
  $m_F=1$ and $m_F=0$ with a significant fraction of atoms in the
  normal cloud (the initial population in $m_F=-1$ is due to
  slight preparation imperfections).
  } \label{f:condensatemagetization}
\end{figure}

In the following we will further concentrate on the thermalization
dominated regime in F=1 $^{87}$Rb and investigate the evolution of
an initially prepared $|1\rangle + |0\rangle $ mixture (see
Fig.~\ref{f:condensatemagetization}).

Due to total spin conservation the only spin dynamics is the
coupling $2|0\rangle \leftrightarrow |1\rangle + |-1\rangle $,
which in this case initially leads to the depletion of the
$|0\rangle $ state in favor of the (initially already populated)
$|1\rangle $ and the (initially empty) $|-1\rangle $ states. As
spin dynamics is mostly occurring in the condensate fraction, low
energy atoms are added to the $|1\rangle $ and $|-1\rangle $
states, which in the case of the $|1\rangle $ state just add to
the condensate fraction (the particle number in the normal
component of this state is saturated for the given temperature).
The case of the newly populated $|-1\rangle $ state is however
significantly different, as this state does not yet possess a
normal component. The low energy atoms in this state quickly (on a
shorter timescale - on the order of 50\,ms than spin dynamics - on
the order of seconds) thermalise with the normal component atoms
of the other spin states. This leads to a slow buildup of the
$|-1\rangle $ normal component and at the same time an increase in
condensate fraction in the $|1\rangle $ component, while the
$|0\rangle $ state condensate fraction decreases.

An interesting point is that this process leads to a slight
decrease in temperature, as it uses energy from the existing
normal components to thermalise the low energy atoms entering the
$|-1\rangle $ state from the $|0\rangle $ condensate fraction.
This temperature decrease at the expense of total condensate
fraction is similar to decoherence driven cooling observed in
quasi spin 1/2 systems~\cite{Lewandowski2003a}.

We want to emphasise that under typical experimental conditions
the thermal energy corresponds to roughly $k_B \times 300\, $nK
and is thus more than an order of magnitude larger than the spin
dependent mean-field shifts of roughly $k_B \times 10\, $nK
responsible for spin dynamics. This directly implies that in
thermal equilibrium the normal components of different spin states
will have equal population (if there are sufficiently many atoms
available in each spin component). For the case discussed in this
paper the $|-1\rangle $ normal component will grow until either
the $|0\rangle $ condensate fraction is completely depleted or it
reaches its saturated occupation for the given temperature, i.e.
the same occupation as the other spin state normal components
(which are saturated, as there exists a condensate fraction in
these components). Only in the second case a condensate fraction
will build up in the $|-1\rangle $ state, which now will be
determined not by the thermal energy scale but by spin
dynamics~\cite{Erhard2004a}.

\begin{figure}
\resizebox{0.45\textwidth}{!}{
  \includegraphics{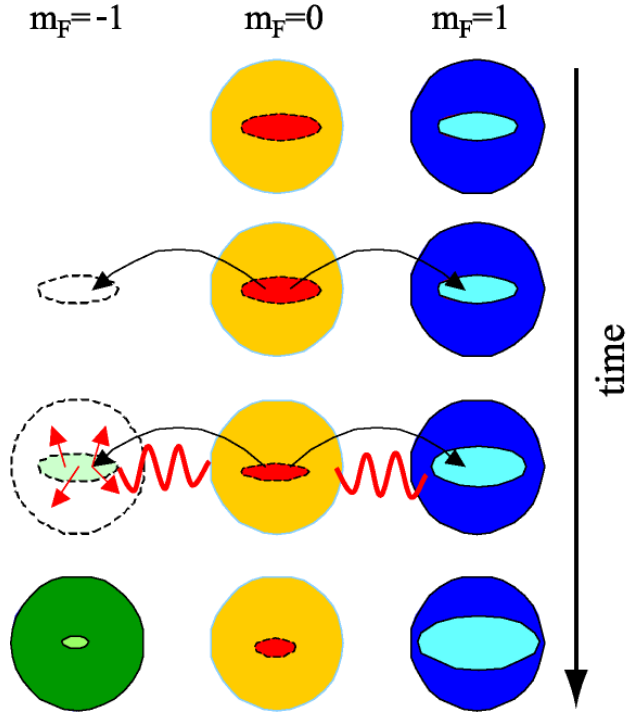}
} \caption{Schematic view of condensate magnetisation in thermally
dominated spin dynamics. The top row shows the initially prepared
state. Spin dynamics is transferring population from the
$|0\rangle $ condensate fraction to the $\pm1 \rangle $ condensate
fractions (second row). Due to fast thermalisation and the absence
of a $|-1\rangle $ normal component, the new $|-1 \rangle $ atoms
start to populate the normal component, while the new $|+1\rangle
$ just add to the corresponding condensate fraction (third row).
In the end there will be thermal equilibrium with equally
populated normal components (zero total spin) and a stronger than
initial condensate fraction magnetisation.}
\label{f:spinormagnetisation}
\end{figure}

In any case this process will tend towards a total zero spin in
the normal components (equal occupation) and thus shift the total
spin of the condensate fractions towards more positive values. The
strongest magnetisation of the condensate fraction occurs if the
initial population of the $|0\rangle $ condensate fraction does
not suffice to saturate the $|-1\rangle $ normal component
occupation via spin dynamics. In this case spin dynamics stops
after the $|0\rangle $ condensate fraction is depleted and only a
$|+1\rangle $ condensate fraction remains, i.e. the condensate
fraction is fully magnetised.

\begin{figure}
\resizebox{0.45\textwidth}{!}{
  \includegraphics{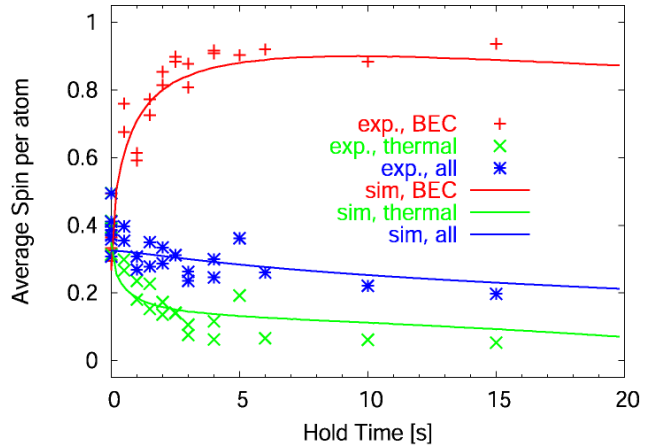}
  }
  \caption{Experimental data versus a simulation for thermally dominated spinor dynamics for an
  initial preparation of the sample in $m_F=1$ and $m_F=0$. The
  graph shows the average magnetisation of atoms in the condensate
  fraction, atoms in the normal fraction and in total.
  } \label{f:BECmagetization}
\end{figure}

The principle mechanisms are again summarized in
Fig.~\ref{f:spinormagnetisation}: The population of the
$|-1\rangle $ condensate part thermalises and fills up a
$|-1\rangle $ normal component. Thus the normal conponent total
spin finally adds up to zero. Due to spin conservation, the
condensate spin has to increase, which is reflected in a higher
$|+1\rangle $ condensate fraction population. This process is
clearly reflected in the experimental data presented in
Fig.~\ref{f:BECmagetization}. The data is well reproduced by a
numerical simulation based on a simple rate equation model,
presented in detail in~\cite{Erhard2004a}. The slight decrease in
the average spin for the total ensemble is due to the fact that
the trap losses are dominated by three body collisions,
predominantly occurring in the magnetised condensate fraction.

In conclusion in this paper we have presented investigations on
spatial variations, influence of magnetic fields and high
temperature as fundamental and new aspects in spinor dynamics. We
found that fast spin dynamics in inhomogeneous (trapped) ensembles
leads to spatial effects which promises new complex coupled
spatial and spin dynamics. We have shown, that the dominant
magnetic field influence stems from the quadratic Zeeman effect,
limiting the offset fields up to which spinor dynamics can be
observed for typical experimental conditions to a few 100\,mG.
Furthermore we investigated the regime of finite temperature
spinor dynamics considering the example of condensate
magnetisation in favour of an equalised spin distribution in the
normal component. This work demonstrates the versatility and
complexity of spinor Bose-Einstein condensates and paves the way
for a broad range of future investigations.

We acknowledge support in SPP1116 of the Deutsche
Forschungsgemeinschaft.

\bibliographystyle{unsrt}
\bibliography{Sengstockxxx}

\end{document}